%% file: main.tex
\documentclass[lettersize,journal,twoside]{IEEEtran}
\IEEEoverridecommandlockouts

\usepackage[pdftex]{graphicx}
\usepackage{amsmath}
\usepackage{algorithmic}
\usepackage{siunitx}
\usepackage{authblk}
\usepackage{booktabs}
\usepackage{multirow}
\usepackage{fancyhdr}
\usepackage{cite}
\usepackage[many]{tcolorbox}
\usepackage{lipsum}
\usepackage{xspace}
\usepackage{RobStd}
\usepackage{colortbl}
\usepackage{xcolor}
\usepackage{amssymb} 
\usepackage{makecell}
\usepackage{tabularx}
\usepackage{siunitx}  

\ifCLASSOPTIONcompsoc
  \usepackage[caption=false,font=normalsize,labelfont=sf,textfont=sf]{subfig}
\else
  \usepackage[caption=false,font=footnotesize]{subfig}
\fi

\input{comments}

\setboolean{showcomments}{false}

\begin{document}
\renewcommand{\arraystretch}{1.1}

\title{A 65 nm Multi-Modal Bayesian Inference Engine with 16.3 fJ/Sample Calibration-Free GRNG for Risk-Aware At-Home Skin Lesion Screening}
 
\author{%
Steven Davis,~\IEEEmembership{Graduate Student Member,~IEEE},~%
Likai Pei,~\IEEEmembership{Graduate Student Member,~IEEE},~%
Jianbo Liu,~\IEEEmembership{Graduate Student Member,~IEEE},~%
Zephan M. Enciso,~\IEEEmembership{Graduate Student Member,~IEEE},~%
Boyang Cheng,~\IEEEmembership{Graduate Student Member,~IEEE},~%
Xueji Zhao,~\IEEEmembership{Graduate Student Member,~IEEE},~%
Danny Z. Chen,~\IEEEmembership{Fellow,~IEEE},~%
Ningyuan Cao,~\IEEEmembership{Member,~IEEE}~%
\thanks{© 2026 IEEE. Personal use of this material is permitted. 
Permission from IEEE must be obtained for all other uses, in any 
current or future media, including reprinting/republishing this 
material for advertising or promotional purposes, creating new 
collective works, for resale or redistribution to servers or lists, 
or reuse of any copyrighted component of this work in other works. 
Published in: \textit{IEEE Transactions on Circuits and Systems I: 
Regular Papers}, DOI: \href{https://doi.org/10.1109/TCSI.2026.3707736}
{10.1109/TCSI.2026.3707736}}
\thanks{The authors are with the College of Engineering at the University of Notre Dame, Notre Dame, IN, USA.}%
}

\markboth{} {}

\maketitle

\begin{abstract}

    We present a 65 \qty{}{\mathbf{\nm}}, risk-aware multi-modal Bayesian inference engine that enables privacy-preserving, fully on-device analysis for safe at-home skin lesion screening under uncontrolled user conditions. The compute-in-memory architecture performs in-word Mixture-of-Gaussian sampling, improving uncertainty fidelity beyond uni-modal Bayesian neural networks (BNNs). This added expressiveness increases equal-risk operating coverage by $\mathbf{1.4\times}$, improves robustness to user-data perturbations by $\mathbf{>1.5\times}$, enhances process variation resilience by $\mathbf{5.5\times}$, and improves balanced accuracy by $\mathbf{1.8\%}$ over state-of-the-art uni-modal BNNs. Hardware robustness is further strengthened by calibration-free Gaussian random number generation using complementary process variation, achieving $\mathbf{16.3}$ \qty{}{\mathbf{\femto\joule}}/sample and $\mathbf{168.6}$ GSa/s/\qty{}{\mathbf{\mm\squared}} efficiency. These results demonstrate a practical, energy-efficient, and risk-aware edge AI solution for privacy-conscious medical screening.
\end{abstract}

\begin{IEEEkeywords}
    Mixture-of-Gaussian; Bayesian Neural Network; Skin Lesion Screening; Imbalanced Dataset; At-Home Screening; Calibration-free; Gaussian Random Number
\end{IEEEkeywords}

\input{00_Introduction}
\input{01_Background}
\input{02_Design}
\input{04_Model}
\input{03_Hardware}
\input{05_Conclusions}

\bibliographystyle{IEEEtran}
\bibliography{references}

\begin{IEEEbiography}[{\includegraphics[width=1in,height=1.25in,clip,keepaspectratio]{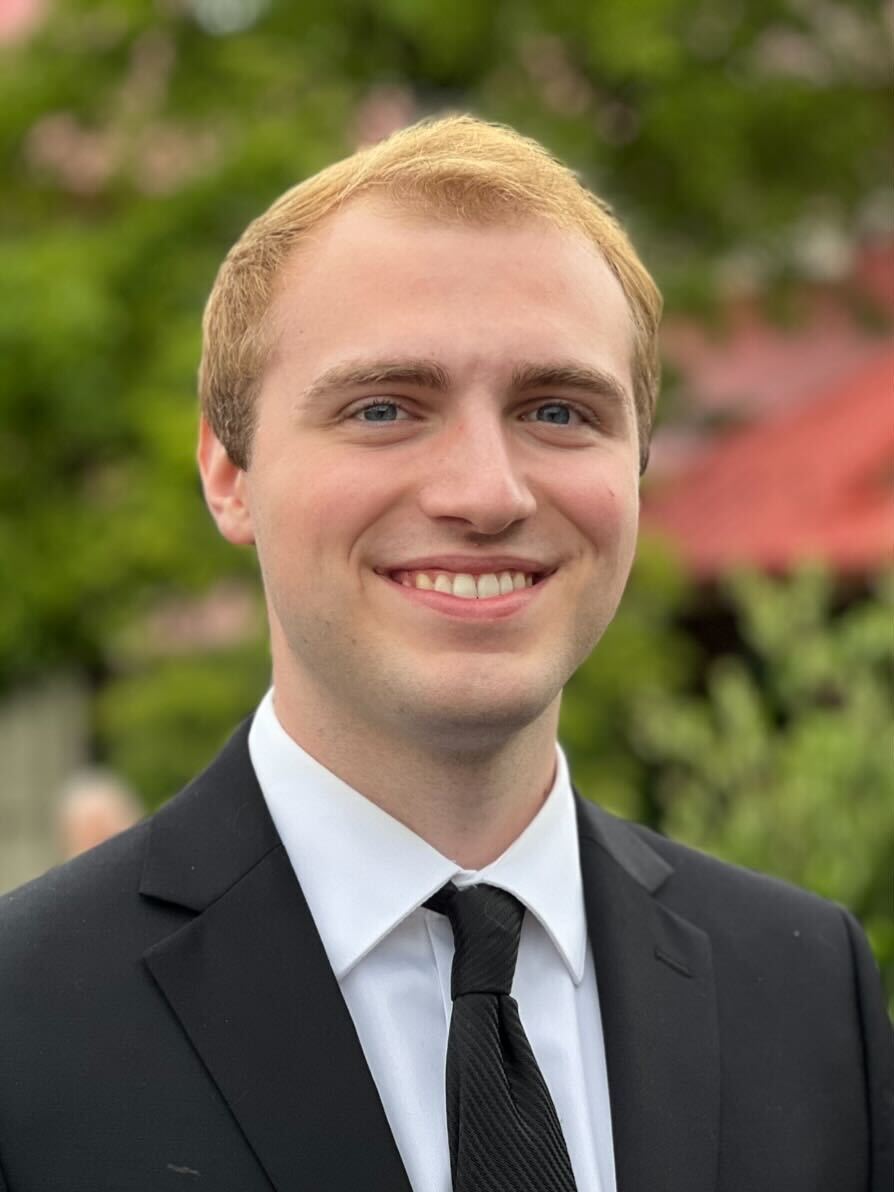}}]{Steven Davis}
received his B.S. degree in Electrical Engineering from The Pennsylvania State University, University Park, PA, USA, in 2022. He is currently pursuing the Ph.D. degree in Electrical Engineering with the University of Notre Dame, Notre Dame, IN, USA.
Since 2022, he joined the Circuit and System Intelligence Research Laboratory, University of Notre Dame. His research interests include analog/mixed-signal circuit design and scalable entropy-centric/reliant machine learning hardware for explainable AI.
\end{IEEEbiography}

\begin{IEEEbiography}
[{\includegraphics[width=1in,height=1.25in,clip,keepaspectratio]{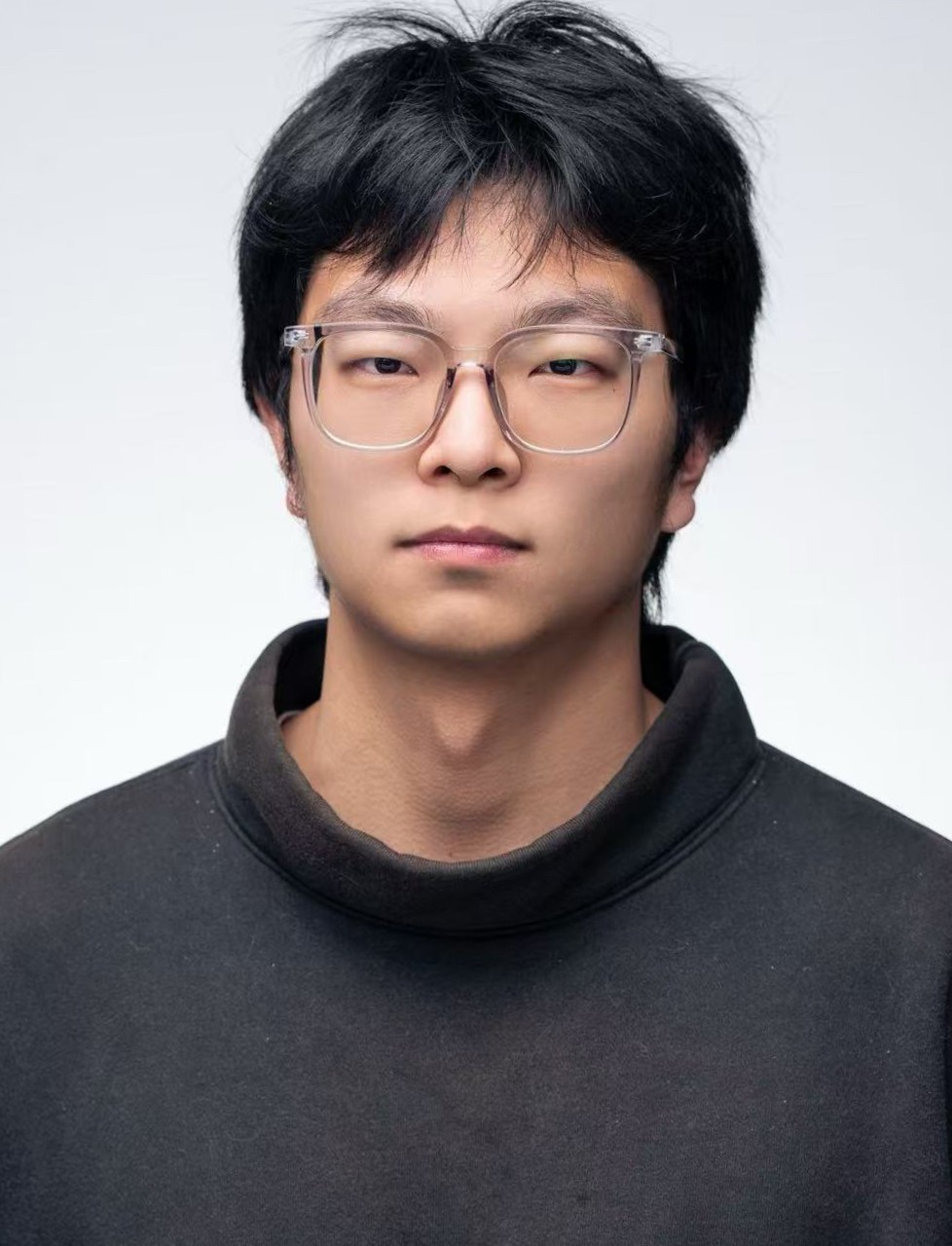}}]{Likai Pei}
received his B.S. degree in Electrical Engineering from Nanjing University of Information Science and Technology, Nanjing, China, in 2023.
He completed his M.S. studies and began pursuing the Ph.D. degree in electrical engineering at the University of Notre Dame, Notre Dame, IN, USA, in 2024.
His research interests include analog/mixed-signal circuit design, explainable artificial intelligence, and low-power neural networks integrated with emerging devices for edge computing.
\end{IEEEbiography}

\begin{IEEEbiography}[{\includegraphics[width=1in,height=1.25in,clip,keepaspectratio]{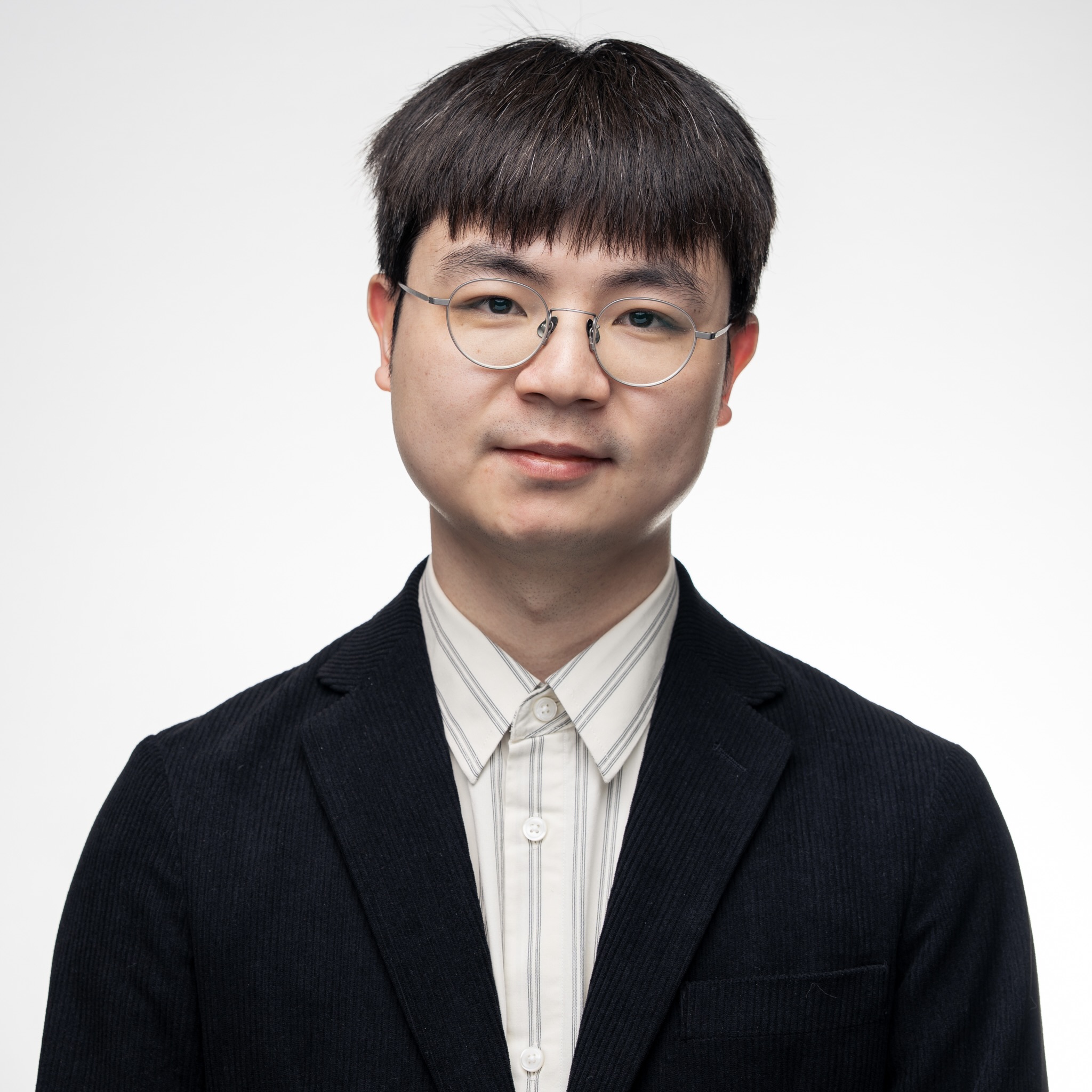}}]{Jianbo Liu}
is currently pursuing the Ph.D. degree in Electrical Engineering with the University of Notre Dame, Notre Dame, IN, USA.
He received his B.S. degree in Electrical Engineering from the Southeast University, Nanjing, China in 2017, and the M.S. degree in Computer Engineering from Northwestern University, Evanston, IL, USA in 2019.

From 2019 to 2021, he was a FPGA Engineer with Ubiquiti Networks, Barrington, IL.
Since 2022, he joined Collaborative AIHW Lab, Notre Dame, IN.
His research interests include privacy in edge devices, and energy-efficient edge system design.

\end{IEEEbiography}

\begin{IEEEbiography}[{\includegraphics[width=1in,height=1.25in,clip,keepaspectratio]{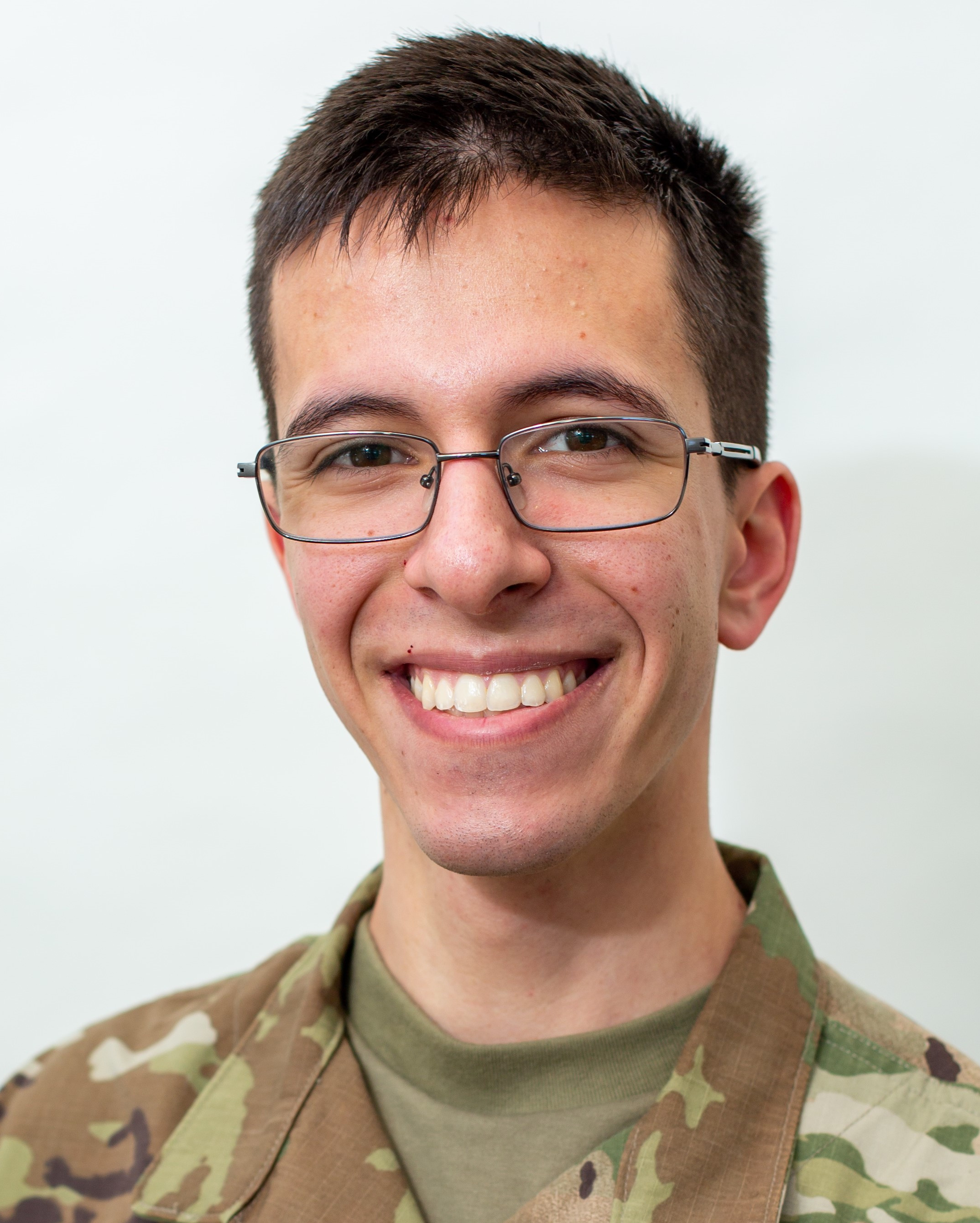}}]{Zephan M. Enciso}
received his B.S. degrees in Computer Engineering and Electrical Engineering from the University of Notre Dame, Notre Dame, Indiana before returning to the University of Notre Dame to pursue his Ph.D.
His research interests include the deployment of edge inference in safety-critical, resource-constrained systems, hardware acceleration of uncertainty-aware artificial intelligence, and novel devices, circuits, and architectures for machine learning.

Z. Enciso was a Department of Defense National Defense Science and Engineering Graduate Fellowship recipient, a Design Automation Conference Young Fellow, and a recipient of the Jack and Mary Ann Remick Fellowship in Engineering.
He was also inducted into Sigma Xi, The Scientific Research Honors Society, the IEEE Eta Kappa Nu Honors Society, and the ACM Upsilon Pi Epsilon Honors Society.
\end{IEEEbiography}

\begin{IEEEbiography}[{\includegraphics[width=1in,height=1.25in,clip,keepaspectratio]{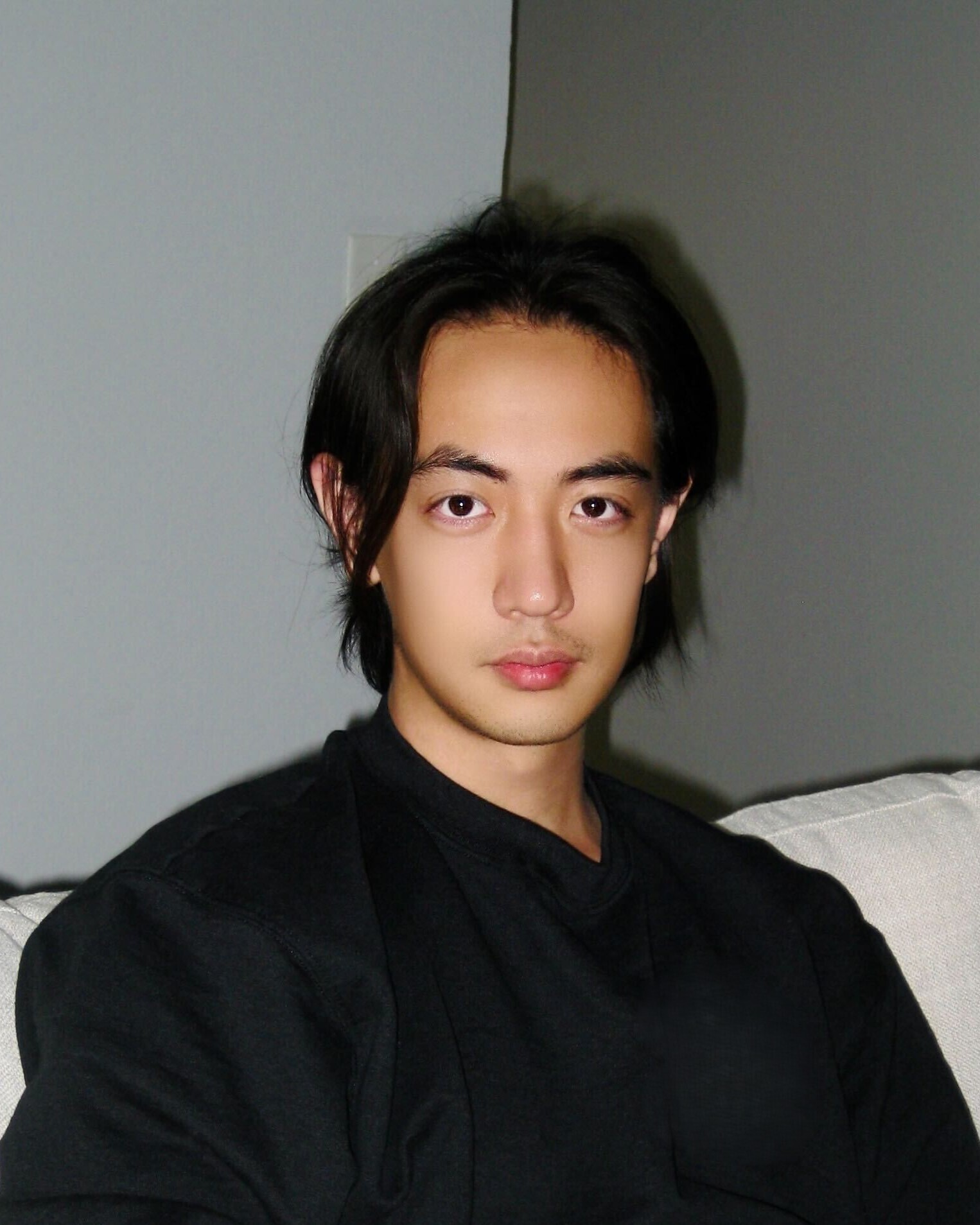}}]{Boyang Cheng}

Boyang received the B.S. degree from Southeast University, Nanjing, China, in 2020. He is currently working toward the Ph.D. degree in Electrical Engineering at the University of Notre Dame. His research interests include digital and analog/mixed-signal circuit design, with emphasis on energy-efficient and explainable/trustworthy hardware systems. 
\end{IEEEbiography}

\begin{IEEEbiography}[{\includegraphics[width=1in,height=1.25in,clip,keepaspectratio]{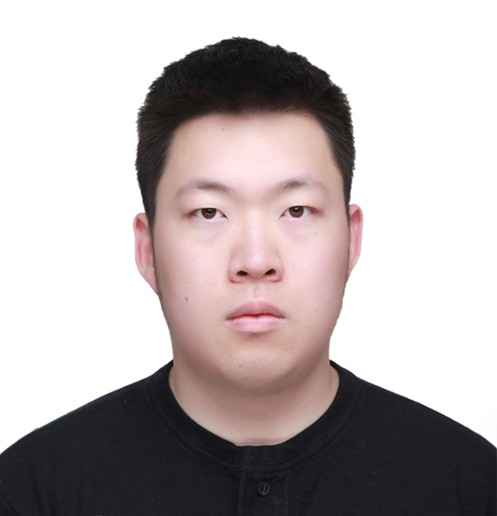}}]{Xueji Zhao}
Xueji Zhao (Student Member, IEEE) received the B.S. degree in electrical and computer engineering from The Ohio State University, Columbus, OH, USA, in 2022, and the M.S. degree in electrical engineering from Columbia University, New York City, NY, USA, in 2023. He is currently pursuing the Ph.D. degree with the University of Notre Dame, Notre Dame, IN, USA, under the supervision of Dr. N. Cao.
His research interests include integrated circuit design and machine learning acceleration.
\end{IEEEbiography}

\begin{IEEEbiography}[{\includegraphics[width=1in,height=1.25in,clip,keepaspectratio]{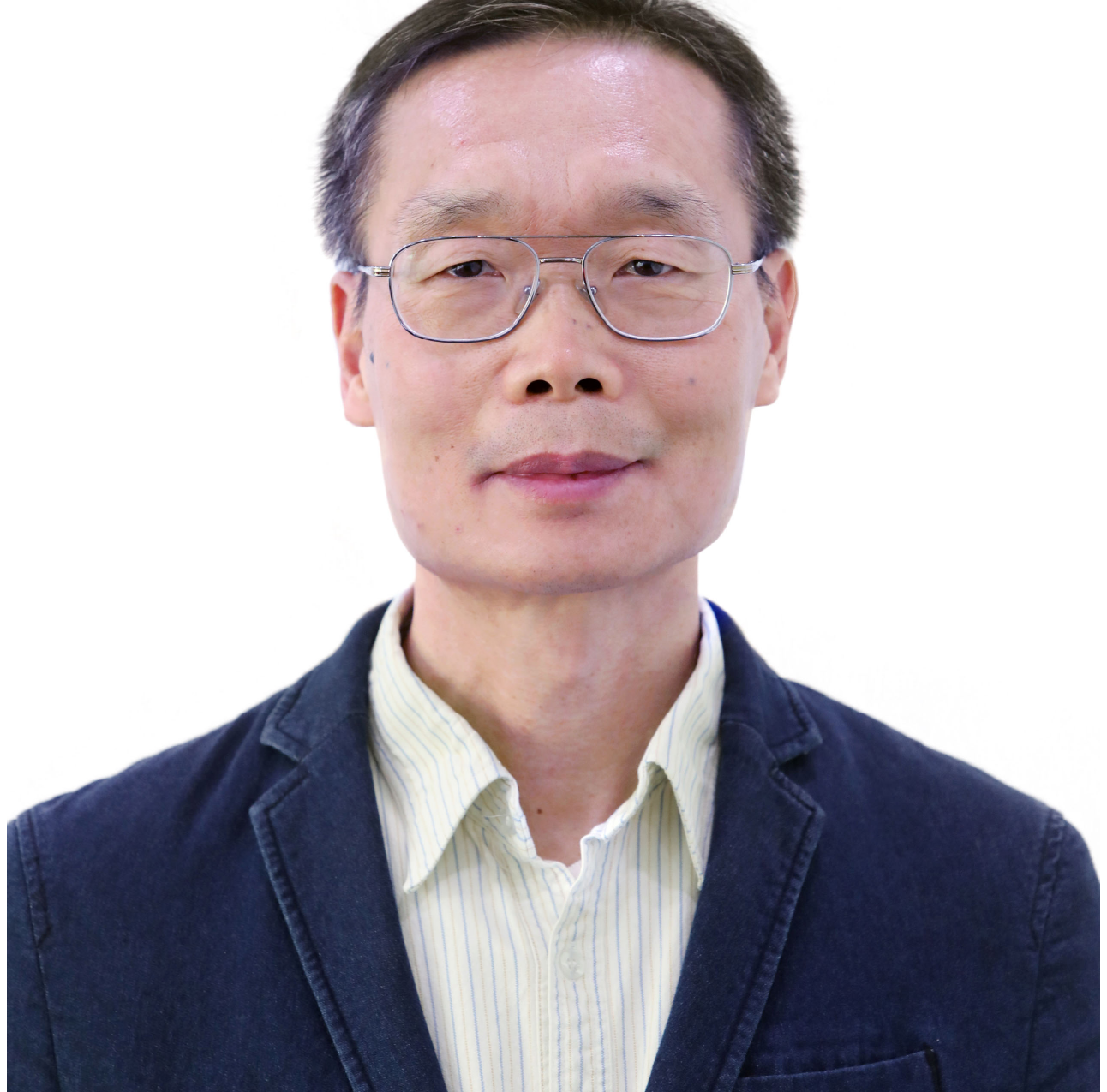}}]{Danny Z. Chen}
Danny Z. Chen received the B.S. degrees in computer science and in
mathematics from the University of San Francisco, CA, USA, in 1985, and the M.S. and Ph.D. degrees in computer science from Purdue University, West Lafayette, Indiana, USA, in 1988 and 1992, respectively. He is currently a Professor with the Department of Computer Science and Engineering, University of Notre Dame, Indiana, USA. His research interests include computational biomedicine, biomedical imaging, computational geometry, algorithms and data structures, machine learning, data mining, and VLSI. He has worked extensively with biomedical researchers and practitioners,
published many papers in these areas, and holds eight U.S. patents for
technology development in biomedical applications. He was the recipient of the US NSF CAREER Award in 1996, Laureate Award in the 2011
Computerworld Honors Program for developing “Arc-Modulated Radiation
Therapy” (a new radiation cancer treatment approach), and the 2017 PNAS
Cozzarelli Prize of the U.S. National Academy of Sciences. He is a Fellow of IEEE and AAAS and a Distinguished Scientist of ACM.
\end{IEEEbiography}

\begin{IEEEbiography}[{\includegraphics[width=1in,height=1.25in,clip,keepaspectratio]{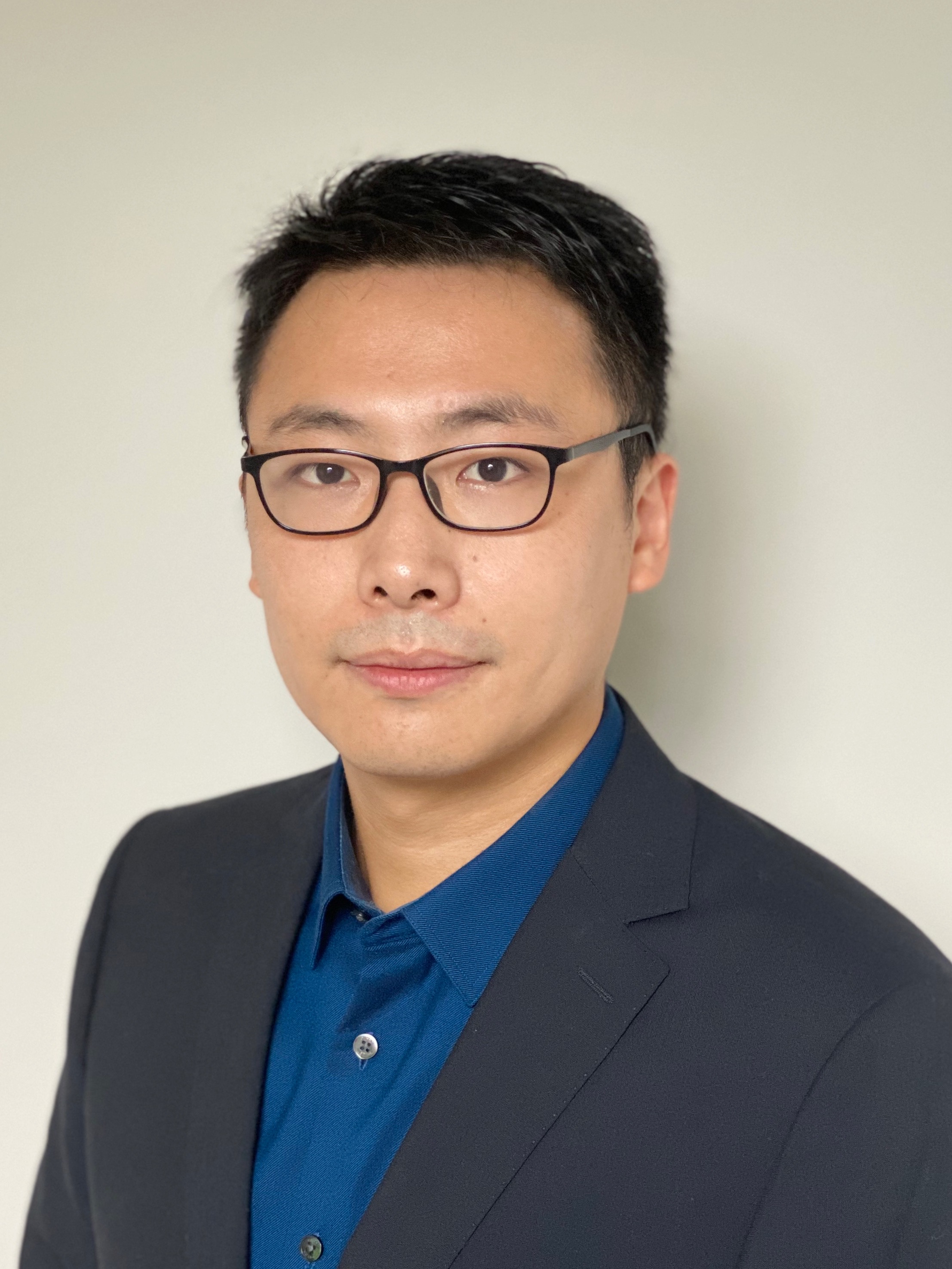}}]{Ningyuan Cao}
Ningyuan Cao (Member, IEEE) received the bachelor’s degree from Shanghai Jiaotong University, Shanghai, China, in 2013, the master’s degree from Columbia University, New York City, NY, USA, in 2015, under the supervision of Dr. Yannis Tsividis, and the Ph.D. degree in integrated circuit and algorithm design for edge intelligence from Georgia Institute of Technology, Atlanta, GA, USA, in 2020, under the supervision of Dr. Arijit Raychowdhury.

He is currently an Assistant Professor with the Department of Electrical Engineering, University of Notre Dame (ND), Notre Dame, IN, USA. Before joining ND, he was a Research Associate with IBM Thomas J. Watson Research Center, Yorktown Heights, NY, USA, for one year. His works have been published/presented/reported by primary journals/conferences/press in various fields of solid-state circuit design, microwave, industrial electronics, and so on (such as ISSCC, IEEE Journal of Solid-State Circuits, IEEE Transactions on Industrial Electronics, IMS, CICC, and IEEE Transactions on Circuits and Systems-I: Regular Papers). His research interests include analog/mixed-signal circuits, digital architecture, and the IoT system design for machine learning acceleration/distributed intelligence; and custom IC design automation with data-driven methods.
\end{IEEEbiography}

\end{document}

%% file: comments.tex
\usepackage{ifthen}
\usepackage{amssymb}
\usepackage{xcolor}

\newboolean{showcomments}
\setboolean{showcomments}{true}

\makeatletter
\newcommand{\mynote}[3]{%
  \ifthenelse{\boolean{showcomments}}{%
   \fbox{\bfseries\sffamily\scriptsize#1}%
   {\small$\blacktriangleright$\textsf{\emph{\color{#3}{#2}}}$\blacktriangleleft$}}%
  {%
   \@bsphack
   \@esphack
  }%
}
\makeatother

%% file: 00_Introduction.tex
\section{Introduction}
\label{sec:intro}

\IEEEPARstart{S}{kin} cancer is among the most common yet most frequently under-detected malignancies, with patient outcomes strongly dependent on early identification~\cite{skin_cancer_1,skin_cancer_2}. Early-stage lesions often exhibit subtle visual features or may evolve slowly, leading to delayed clinical evaluation and more costly medical procedures~\cite{diagnostic_delays,skin_cancer_costs}. Skin cancer represents approximately one-third of all cancer diagnoses worldwide, with an estimated one in five Americans expected to be diagnosed during their lifetime~\cite{WHO_stats}. Although highly treatable when detected early, recent U.S. estimates report over 100,000 new melanoma cases and more than 10,000 deaths annually~\cite{sc_stats}. Moreover, patients with a history of malignant lesions, as well as individuals with pre-cancerous abnormalities, remain susceptible to recurrence or the emergence of new high-risk lesions, necessitating continual surveillance even after initial treatment or diagnosis~\cite{recurrence, emergence}. These trends highlight a persistent gap between clinical detectability and real-world screening, motivating the need for frequent, accessible, and reliable monitoring beyond traditional in-clinic workflows.

\begin{figure}
    \centering
    \includegraphics[width=\columnwidth]{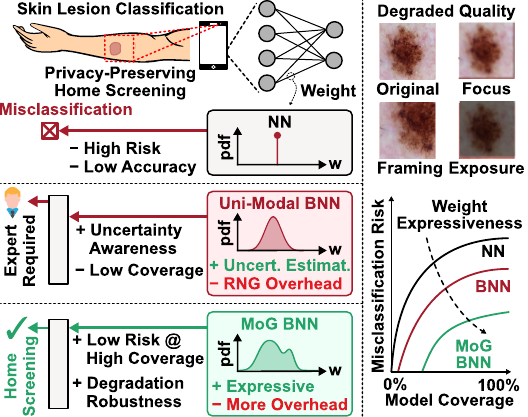}
    \caption{At-home medical screening offers rapid assessment and privacy, but suffers from user-error and unreliable environments. MoG BNNs reduce the risk associated with misclassifications.}
    \label{fig:motivation}
    \vspace{-0.5cm}
\end{figure}

\IEEEpubidadjcol

Protecting patient privacy and sensitive medical data is essential for enabling frequent, user-driven screening. Medical monitoring technologies now support rapid, at-home disease assessment without requiring clinical visits, allowing earlier intervention and longitudinal observation~\cite{remote_monitoring,IoMT}. However, many existing solutions remain dependent on healthcare-provider mediation or cloud-based analytics, limiting autonomous operation. Reliance on centralized computing also introduces additional risks, including exposure of medical records~\cite{privacy_concerns1,privacy_concerns2} and susceptibility to adversarial or data-poisoning attacks on shared datasets and models~\cite{adversarial,data_poisoning}. By contrast, fully on-device screening performs inference locally, reducing communication latency and data exposure—properties particularly important for recurring tasks such as skin lesion monitoring~\cite{edge_IoMT}. Accordingly, this work defines privacy as executing screening and inference entirely on-device, without transmitting sensitive medical data to external servers or shared computational infrastructure.

However, at-home screening operates under uncontrolled acquisition conditions, including user handling variability, lighting changes, motion blur, and focus misalignment~\cite{PGHD,home_medical_devices} (Fig.~\ref{fig:motivation}). Medical datasets are also inherently imbalanced, with rare but high risk cases underrepresented, and deployment conditions often differ from training due to evolving users and environments~\cite{ISIC1,ISIC2}. Together, these factors increase the risk of overconfident false-negative (missed cancerous) predictions, which are unacceptable in life-threatening settings.

Deterministic neural networks are not explicitly designed to quantify predictive uncertainty and commonly rely on softmax scores, which provide relative class confidence rather than uncertainty about the model’s predictions and can be sensitive to noise~\cite{Uncertainty_1}. Bayesian neural networks (BNNs) address this limitation by modeling weights probabilistically, enabling calibrated uncertainty estimation and risk-aware inference~\cite{BNN_1,BNN_2}. However, most BNN implementations assume uni-modal Gaussian weight distributions, which limits expressiveness and degrades uncertainty fidelity~\cite{Uni_modal,MoG}.

Mixture-of-Gaussian (MoG) BNNs provide a natural extension by representing each weight as a mixture of $K$ Gaussian components, enabling approximation of complex distribution posteriors and improved uncertainty fidelity~\cite{MoG,CBLN}. Thus, this expressiveness enhances the risk–model coverage tradeoffs, especially under the variability of user inputs (Fig.~\ref{fig:motivation}). Despite these advantages, MoG BNNs are impractical on existing edge hardware due to increased storage, sampling complexity, and poor scaling with $K$.

Recent BNN accelerators~\cite{liu_ISSCC25,jssc_enciso,intel_JSSC23,STT_ISSCC25,pei_ICCAD24,pei_dac25} integrate Gaussian random number generators (GRNGs) with compute-in-memory (CIM) architectures to reduce sampling overhead. Nevertheless, state-of-the-art (SOTA) CIM-based designs remain restricted to uni-modal distributions and face two key limitations: (1) near-CIM GRNGs incur data movement overhead that undermines in-memory efficiency~\cite{intel_JSSC23,STT_ISSCC25}, and (2) in-word analog GRNGs suffer from device-to-device (D2D) variation, requiring costly and fragile calibration—particularly under near- and sub-threshold operation~\cite{liu_ISSCC25,jssc_enciso,pei_dac25}. These constraints prevent scalable support for expressive probabilistic models.

This work, to the best of our knowledge, presents the first CIM Bayesian inference engine that supports scalable, calibration-free MoG sampling. This work makes the following key contributions:

\begin{itemize}
    \item \textbf{First In-Memory MoG BNN Hardware}: We realize a CIM system that performs MoG weight sampling with a programmable mixture order ($K = 1$–$16$), enabling post-fabrication tuning of model expressiveness versus throughput, and demonstrating up to 337.3~GOp/s/\qty{}{\mm\squared}.
    
    \item \textbf{Calibration-Free In-Word GRNG}: A novel GRNG exploits complementary process variation for ultra-low-energy (16.3~\qty{}{\femto\joule}/sample) operation, while eliminating D2D offset calibration, achieving 168.6~GSa/s/\qty{}{\mm\squared} sampling.

    \item \textbf{Robust At-Home Skin Lesion Screening}: MoG BNN models demonstrate improved class-balanced accuracy (1.8\%), expanded model coverage at equal-risk (1.4$\times$), and enhanced resilience to at-home screening-induced user error ($>$1.5$\times$) compared to SOTA uni-modal BNNs.
\end{itemize}

%% file: 01_Background.tex
\section{Background}
\label{sec:background}

\subsection{Deep Learning for Skin Lesion Classification}

Deep learning, particularly convolutional neural networks (CNNs), has become the dominant approach for automated skin lesion classification, with diagnostic performance strongly influenced by network architecture, feature extractor choice, and training protocol. Comparative studies of modern backbones—including VGG, Inception, DenseNet, and ResNet variants—report substantial variability in balanced accuracy and per-class recall on benchmarks from challenges hosted by the International Skin Imaging Collaboration (ISIC)~\cite{ISIC_challenge}, even when models are trained on identical datasets~\cite{ISIC1,ISIC2}. These results highlight the sensitivity of skin lesion classification performance to architectural design choices rather than the existence of a single universally optimal model~\cite{ISIC_challenge, Menegola_2017}. Despite this variability, CNN-based systems have consistently achieved performance comparable to or exceeding that of dermatologists for melanoma detection and benign--malignant discrimination~\cite{CNN_v_Man, CNN_v_Man2}. 

As a result, most contemporary skin lesion classification architectures have emphasized robustness through architecture selection, transfer or ensemble learning, and imbalance-aware training strategies; however, they remain primarily optimized for predictive performance under balanced evaluation metrics and controlled evaluation settings, with comparatively less emphasis on expressive weight representation for minority class modes or uncertainty quantification with risk-aware decision-making. This gap becomes particularly important for at-home skin lesion screening scenarios, where image quality, user variability, and ambiguous cases are unavoidable and where actionable risk awareness is essential. Consequently, the integration of calibrated uncertainty estimation and risk-sensitive inference into real-time deployment remains an open challenge, motivating alternative probabilistic modeling approaches and hardware-efficient implementations.

\subsection{Bayesian Neural Networks}
\subsubsection{BNN Overview}
\label{sec:bnn_overview}
BNNs are distinguished by their ability to provide  probabilistic estimates of uncertainty with their predictions~\cite{BNN_1,BNN_2}. This capability is essential for deployments that require decision robustness, especially when inference data diverges significantly from training data.

The core mathematical principle of BNNs is encapsulated in the posterior distribution of the weights, formalized as:
\begin{equation}
\begin{aligned}
P(\mathbf{W} | \mathbf{X}, \mathbf{Y}) = \frac{P\left(\mathbf{Y} | \mathbf{W}, \mathbf{X}\right) P(\mathbf{W})}{P(\mathbf{X})},
\end{aligned}
\label{eq:posterior}
\end{equation}

\noindent where \(P(\mathbf{W} | \mathbf{X}, \mathbf{Y})\) denotes the posterior probability of the weights \(\mathbf{W}\) given the input \(\mathbf{X}\) and output \(\mathbf{Y}\), \(P(\mathbf{Y} | \mathbf{W}, \mathbf{X})\) represents the likelihood of observing the given outputs for a set of weights and inputs, \(P(\mathbf{W})\) is the prior distribution of the weights, and \(P(\mathbf{X})\) is the evidence or the probability of observing the input data.

The computational complexity of directly calculating the posterior distribution in BNNs for each weight often requires variational inference methods. These methods in BNNs approximate the posterior of the weights using a Gaussian distribution $w \sim \mathcal{N}(\mu, \sigma)$, where $\mu$ and $\sigma$ represent the mean and standard deviation variational parameters~\cite{Uni_modal,MoG}, which are trained for each weight by employing Bayes by Backprop~\cite{Bayes_by}. 

\subsubsection{BNN Hardware}
BNN inference requires multiple samples per input to estimate the mean and variance of the classification score, which are essential for quantifying uncertainty. In resource-constrained environments, such as IoT and edge computing, energy-efficient custom processors or CIM units are critical to optimize inference.

Recent BNN accelerators~\cite{liu_ISSCC25,jssc_enciso,intel_JSSC23,STT_ISSCC25,pei_ICCAD24,pei_dac25} enhance hardware efficiency through optimized GRNG techniques and leverage CIM architectures to reduce memory bottlenecks. However, these solutions focus on efficient GRNG integration, neglecting the complexities of sampling calibration and managing multi-distribution storage and generation.

SOTA accelerators also suggest further simplification of weight generation through dynamic weight decomposition, in which weights are alternatively expressed as \( w = \mu + \epsilon \cdot \sigma \), where \(\epsilon\) is drawn from a standard normal distribution \(\epsilon \sim \mathcal{N}(0,1)\), which presents insights for the storage of decomposed distributions within a CIM hardware architecture~\cite{liu_ISSCC25,jssc_enciso,pei_ICCAD24,pei_dac25}. 

\subsection{Mixture-of-Gaussian (MoG) BNN}
\label{sec:gmm}

MoG weight representations provide an expressive Bayesian parameterization in which each weight is modeled as a multi-modal variable rather than a single uni-modal distribution. By expressing each parameter as a finite mixture-of-Gaussian components, these formulations can approximate arbitrarily complex weight distributions, with larger values of $K$ enabling increasingly rich representations~\cite{MoG,GMM}. This multi-modal structure preserves multiple plausible hypotheses within the weight space, allowing MoG-based BNNs to more faithfully capture epistemic uncertainty arising from ambiguous inputs, class imbalance, or limited training data. In contrast to Gaussian addition or ensemble-based approaches, which aggregate uncertainty through linear superposition or prediction averaging, MoG weights represent uncertainty as an explicit categorical mixture over distinct parameter modes, introducing a higher-order stochastic structure that cannot be captured by single-Gaussian or averaged representations~\cite{CBLN,MoG}.

Statistically, each MoG weight is encoded as a mixture of $K$ uni-modal Gaussian components whose contributions are governed by mixing ratios ($\pi$) forming a categorical distribution. The $k$-th component is parameterized by its mean and variance, while its associated coefficient $\pi_k,\ k = 1, 2, \ldots, K$, defines the probability of selecting that component during inference under the constraint $\sum_{k=1}^{K}\pi_k = 1$. This representation leverages Gaussian mixture modeling to approximate complex Bayesian weight distributions~\cite{GMM}. The coefficients are initialized uniformly to $\frac{1}{K}$ and refined using the expectation--maximization algorithm to obtain the likelihood of each mode~\cite{CBLN,GMM}, allowing the model to emphasize dominant modes while still preserving minority hypotheses.

\begin{figure}
    \centering
    \includegraphics[width=\columnwidth]{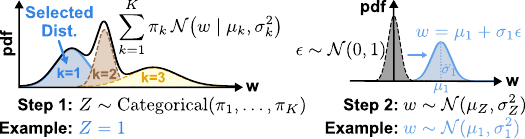}
    \caption{MoG Sampling requires two steps: (1) Categorically selecting one of $K$ distributions to sample from based on the mixing ratios and (2) scaling a standard normal distribution ($\epsilon$) to the selected parameters.}
    \label{fig:MoG_samp}
    \vspace{-0.3cm}
\end{figure}

As illustrated in Figure~\ref{fig:MoG_samp}, sampling from a trained MoG weight distribution proceeds in two steps. First, a latent categorical variable $Z$ is drawn according to the learned mixing ratios to select one of the $K$ Gaussian components:

\vspace{-0.2cm}
\begin{equation}
Z \sim \text{Categorical}(\pi_1, \dots, \pi_{K})
\end{equation}
\vspace{-0.35cm}

For example, when $K=3$, each mode has a $\pi_k,\ k = 1, 2, 3$ probability of being selected per inference. Second, a standard normal sample $\epsilon \sim \mathcal{N}(0,1)$ is transformed via weight decomposition using the mean and variance of the selected component, yielding a sample from the chosen uni-modal distribution:

\vspace{-0.2cm}
\begin{equation}
w = \mu_Z + \sigma_Z \epsilon, \quad \epsilon \sim \mathcal{N}(0, 1)
\end{equation}
\vspace{-0.35cm}

Over repeated inference passes, where the selected weight component is independently sampled at each iteration, the aggregate behavior recovers the full multi-modal distribution.

%% file: 02_Design.tex
\section{Architecture and Circuit Design}
\label{sec:design}

\subsection{System Overview}

\begin{figure}[b!]
    \centering
    \vspace{-0.3cm}
    \includegraphics[width=\columnwidth]{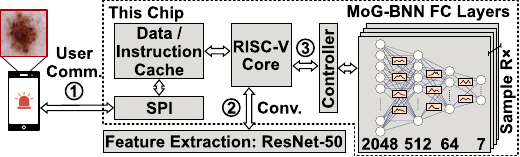}
    \caption{The system level overview of on-device skin-lesion feature extraction and MoG Bayesian inference.}
    \label{fig:system}
\end{figure}

Figure~\ref{fig:system} illustrates the system-level architecture of this work's skin lesion classification engine. An on-device RISC-V core coordinates data movement between a standard convolutional feature extraction backbone and this work’s Bayesian fully connected (FC) inference engine. The three FC layers utilize in-memory MoG weights, and each input is stochastically sampled $R\times$, with the resulting output inference variance providing the uncertainty quantification. This design leverages established algorithmic findings indicating that restricting stochasticity to the FC layers resolves overconfidence and achieves uncertainty awareness comparable with full-network Bayesian models~\cite{last_layer3, last_layer1, last_layer2}. By avoiding the prohibitive computational overhead of stochastic convolutions, the design focuses on the FC layers, which represent the primary bottleneck due to their low arithmetic intensity. The high data movement and minimal data reuse inherent to these layers make them ideal candidates for in-memory computation~\cite{Han1}. Consequently, improved expressiveness in this design specifically refers to increasing the number of Gaussian components stored within the FC weight distributions.

\begin{figure}
    \centering
    \includegraphics[width=\columnwidth]{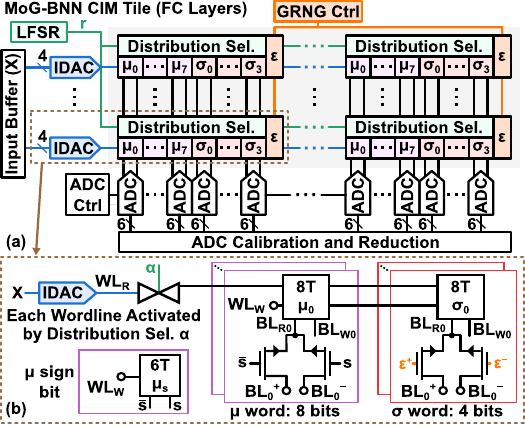}
    \caption{(a) The MoG BNN FC layer CIM tile footprint. (b) Single word architecture with $\mu$ and $\sigma$ SRAM storage and bitline discharge control from the MoG distribution selection ($\alpha$) and GRNG ($\epsilon$).}
    \label{fig:arch}
    \vspace{-0.3cm}
\end{figure}

Figure~\ref{fig:arch}(a) further showcases the footprint of the CIM MoG FC layers. Within each word, weights are decomposed and stored as an 8-bit signed mean ($\mu$) and a 4-bit unsigned standard deviation ($\sigma$). In-word GRNGs generate the standard normal $\epsilon$ sample. This design is consistent with SOTA weight decomposition~\cite{liu_ISSCC25}, with the exception that $\mu$ and $\sigma$ are stored together rather than individual tiles, in order to enable MoG shaping. In-word distribution selectors achieve this by aggregating neighboring words and only enable the sampling of one of $K$ grouped words per inference. During matrix–vector multiplication (MVM), 4-bit inputs drive current digital-to-analog converters (IDACs), while columns are digitized by 6-bit successive approximation register analog-to-digital converters (ADCs). ADCs are pitch-matched to the words to optimize area efficiency and enable single-cycle MVM.

\subsection{CIM Architecture}

\subsubsection{Weight Storage}

Both $\mu$ and $\sigma$ parameters are stored using 8T SRAM cells to improve data isolation and reduce leakage and parasitic coupling during analog operation (Fig.~\ref{fig:arch}(b)). The sign bit of $\mu$ ($\mu_s$) is stored in a compact 6T cell. During $X\mu$ computation, $\mu_s$ steers bitline discharge toward either BL$^+$ or BL$^-$, implementing signed accumulation. During $X\sigma\epsilon$, the signed Gaussian sample $\epsilon$ controls the discharge duration on BL$^\pm$, with the pulse width representing a standard normal sample scaled by $\sigma$.

The key difference from previous in-memory BNN computation is the distribution selector's output, $\alpha$, gates the word-line voltage from the IDAC, enabling bitline discharge only when the corresponding parameters are categorically selected. This approach allows MoG expressiveness to scale without modifying the underlying memory array.

\subsubsection{Distribution Selection}

\begin{figure}
    \centering
    \includegraphics[width=\columnwidth]{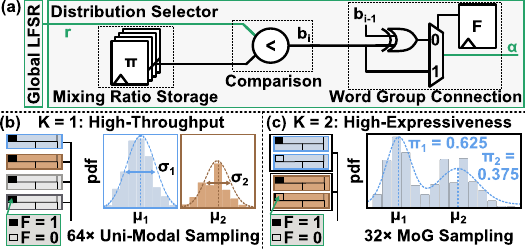}
    \caption{(a) Distribution selector circuit architecture with mixing ratio ($\pi$) storage and adjacent word daisy-chaining ($F$). Example configuration of (b) $K=1$ word storage and (c) $K=2$ word groups, representing the capabilities of post-fabrication $K$ selection and the throughput vs. expressiveness tradeoff.}
    \label{fig:MoG}
    \vspace{-0.3cm}
\end{figure}

Figure~\ref{fig:MoG}(a) shows the in-word distribution selector used to aggregate adjacent words and implement categorical sampling among $K$ Gaussian components by generating $\alpha$ for each word. To implement this sampler in hardware, for each Gaussian component, a 4-bit mixing ratio $\pi_i$ is encoded in local D flip-flops. During each sampling iteration, the cumulative mixing ratio is compared against a 4-bit uniform random value $r$ generated by a global 12-bit linear feedback shift register (LFSR) shared across the tile. This comparison yields a binary intermediate signal $b_i$ for $i \in \{1, \dots, K\}$:

\vspace{-0.15cm}
\begin{equation}
b_i = \begin{cases} 1, & \text{if } r \leq \sum_{n=1}^{i} \pi_n \\ 0, & \text{otherwise} \end{cases}
\end{equation}
\vspace{-0.15cm}

The comparison result determines whether the corresponding word is selected for the MAC operation during that iteration. To enable scalable MoG formation, a single-bit flag ($F$) daisy-chains adjacent selectors to aggregating words groups of $K$ parameters, enabling programmable mixture order ($K = 1$–$16$) post-fabrication. This allows a tunable tradeoff between expressiveness and area-normalized throughput (Fig.~\ref{fig:MoG}(b,c)).

The flag bit functions as follows: when $F_i=1$, the selector marks the start of a new word group (i.e., $i = 0$), thus the comparison result $b_i$ serves directly as the selector's output ($\alpha_i$). When $F_i=0$, $b_i$ is XORed with the previous word’s comparison result ($b_{i-1}$), ensuring that exactly one of the $K$ grouped components is active per sample to prevent simultaneous multi-mode accumulation. This programmable $K$ logic and the final selector output $\alpha_i$ can be explicitly formulated as:

\vspace{-0.15cm}
\begin{equation}
\alpha_i = \begin{cases} b_i, & \text{if } F_i = 1 \\ b_i \oplus b_{i-1}, & \text{if } F_i = 0 \end{cases}
\end{equation}
\vspace{-0.15cm}

\noindent yielding the final aggregated sampled weight for the group:

\vspace{-0.15cm}
\begin{equation}
w = \sum_{i=1}^{K} \alpha_i (\mu_i + \sigma_i \epsilon)
\end{equation}
\vspace{-0.15cm}

Using a single global LFSR per tile enables statistically correct MoG sampling over repeated iterations while eliminating the need for costly per-word or independent random sources typically assumed in MoG and GMM implementations~\cite{CBLN,MoG,GMM}. Although both local and global distribution selection provide equivalent MoG shaping—and thus comparable expressiveness for risk-aware inference—the global LFSR achieves this with substantially lower hardware overhead by sharing randomness across the array, reducing LFSR area overhead proportionally to the tile size. More importantly, global selection consistently improves accuracy on both ISIC 2018~\cite{ISIC1,ISIC2} (see Section~\ref{sec:model}) and CIFAR-10 (Table~\ref{tab:global_lfsr}). We hypothesize that this improvement arises because global distribution selection preserves coherent interactions among neighboring weights as each mixture component is trained independently as a complete uni-modal model and combined into a MoG distribution at inference time. In contrast, local in-word selection mixes components at a finer granularity, breaking these learned correlations and degrading accuracy.

\begin{table}[t!]
\centering
\caption{Balanced accuracy with local $r$ versus global $r$ generation for MoG distribution selection.}
\vspace{-0.1cm}
\label{tab:global_lfsr}
\renewcommand{\arraystretch}{1.15}
\begin{tabular*}{\linewidth}{@{\extracolsep{\fill}}lcc}
\hline
\textbf{Dataset} & \textbf{Local $r$} & \textbf{Global $r$} \\
\hline
ISIC 2018$^\dagger$  & 72.41\%  & \textbf{74.8\%} \\
CIFAR-10$^\ddagger$ & 84.06\% & \textbf{88.1\%} \\
\hline
\end{tabular*}
\vspace{2pt}

\begin{minipage}{\linewidth}
\raggedright
$^\dagger$ResNet-50 Feature Extractor \\
$^\ddagger$MobileNet Feature Extractor
\end{minipage}
\vspace{-0.6cm}
\end{table}

\subsection{In-Word Calibration-Free GRNG}

\subsubsection{SOTA GRNG Offset and Calibration Overhead}

\begin{figure}[b!]
    \centering
    \vspace{-0.3cm}
    \includegraphics[width=\columnwidth]{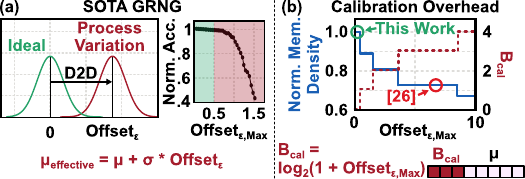}
    \caption{(a) SOTA in-word GRNG hardware suffers from device-to-device (D2D) process variation that impacts the weight's effective mean. (b) This requires costly calibration that requires additional calibration bits ($B_{cal}$), significantly degrading memory density.}
    \label{fig:grng_cal}

\end{figure}

In-memory Gaussian random number generators (GRNGs) are a critical bottleneck in BNN accelerators, as sampling energy and throughput directly impact overall system efficiency. Prior in-word GRNGs primarily rely on dynamic entropy sources such as thermal noise or tunneling effects~\cite{liu_ISSCC25,jssc_enciso,pei_dac25}. While dynamically random, these entropy sources are subtle in scaled CMOS technologies. As a result, D2D variation introduces static offsets (Offset$_\epsilon$), shifting the output mean away from zero (Fig.~\ref{fig:grng_cal}(a)). Simulations on the skin lesion screening data (see Section~\ref{sec:model}) show that an offset comparable to the size of the cycle-to-cycle standard deviation degrades the accuracy by up to $14\%$ and quickly drops off thereafter. This offset can be calibrated through the stored $\mu$ parameter; however, this requires a costly calibration cycle and additional bits ($B_{cal}$), significantly reducing the memory density (Fig.~\ref{fig:grng_cal}(b)).

\subsubsection{Calibration-Free Process Variation-based Operating Principles}

\begin{figure}[t!]
    \centering
    \includegraphics[width=\columnwidth]{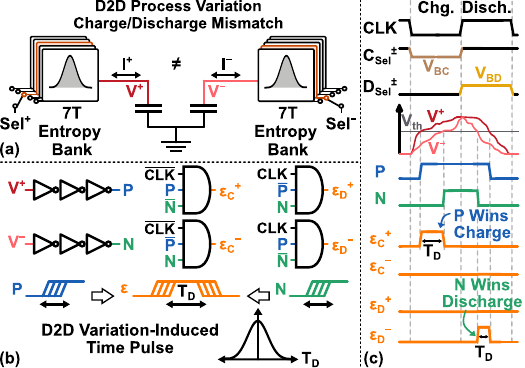}
    \caption{(a) This work eliminates calibration by leveraging D2D variation by randomly selecting a device from an entropy bank and extracting the capacitor charge/discharge mismatch of 2 identical setups. (b) The capacitor voltages are sharpened and the difference is converted to a time pulse that follows a standard normal distribution. (c) Example GRNG timing diagram.}
    \label{fig:grng_op}
    \vspace{-0.3cm}
\end{figure}

Figure~\ref{fig:grng_op}(a) illustrates the operating principle of this work's GRNG. Rather than suppressing or calibrating process variation, this work exploits it as the primary entropy source. By intentionally extracting complementary D2D differences between randomly selected transistors, the GRNG produces zero-mean Gaussian samples in a calibration-free manner. In this context, ``calibration-free'' strictly refers to the elimination of the per-device tuning required to compensate for static, device-specific offsets in traditional dynamic noise sources based on the calibration process described in~\cite{liu_ISSCC25,jssc_enciso}.

For each sample, a global LFSR (shared with the distribution selector and every GRNG cell) randomly selects one device from two identical entropy banks, each consisting of seven minimum-sized transistors. Under random selection, the difference between two independent, identically distributed device parameters—such as threshold voltage ($V_{th}$) variations or drive current—follows a zero-mean Gaussian distribution. The sampled output is extracted from the complementary difference in charging/discharging identical 1 fF capacitors. These are physically implemented as metal fringe capacitors, which are placed directly above the GRNG circuit for optimal area utilization and mismatch performance~\cite{metal_fringe}. 

The capacitor voltages are monitored by inverters, which sharpen the transitions as they cross the switching threshold, generating positive (P) and negative (N) edge signals. The timing difference between these edges is extracted using simple logic, producing signed pulses whose widths ($T_D$) represent samples from a zero-mean Gaussian distribution (Fig.~\ref{fig:grng_op}(b)).

\begin{figure}[b!]
    \centering
    \vspace{-0.3cm}
    \includegraphics[width=\columnwidth]{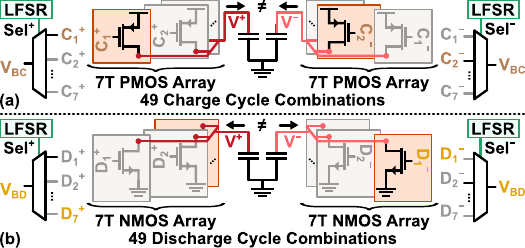}
    \caption{(a) Charge cycle: $V_{BC}$ is delivered to two randomly selected PMOS devices from the two 7T arrays and the capacitor charge time difference is extracted. (b) Discharge cycle: $V_{BD}$ is delivered to two randomly selected NMOS devices to extract the discharge difference. Each cycle yields 49 distinct transistor pairings.}
    \label{fig:grng}
\end{figure}

\subsubsection{Circuit-Level and Dual-Edge Sampling}

This design further leverages a complementary structure by enabling dual-edge operation, for a charge and discharge cycle. Figure~\ref{fig:grng} shows the circuit details used to activate the charge/discharge cycles of the capacitors. During the charging phase, the selected PMOS devices—biased by a tuned voltage $V_{BC}$—inject current into the capacitors, while all unselected devices remain off and do not affect the charge rate (Fig.~\ref{fig:grng}(a)). Due to D2D variation, one capacitor charges faster than the other, producing the differential timing offset. During the discharging phase, an analogous operation is performed using NMOS devices biased by the tuned voltage $V_{BD}$ (Fig.~\ref{fig:grng}(b)). Since process variation is a static phenomenon, this configuration yields 49 distinct transistor pairings for both the charge (PMOS) and discharge (NMOS) cycles. Exploiting both PMOS charging and NMOS discharging thus enables dual-edge operation, effectively doubling GRNG throughput with minimal area overhead. With two independent samples per clock cycle, each GRNG is shared between two $\sigma$ words (one receiving charge cycle pulses and the other discharge cycle pulses, as depicted with the four $\epsilon$ traces in Fig.~\ref{fig:grng_op}(b)), effectively converting a single GRNG circuit into two GRNG cells. Figure~\ref{fig:grng_op}(c) shows an example timing diagram for a full clock cycle.

%% file: 04_Model.tex
\section{Model Evaluation}
\label{sec:model}

The hardware system is evaluated on the ISIC 2018 Task~3 skin lesion classification challenge dataset~\cite{ISIC1,ISIC2}, a seven-class benchmark exhibiting severe real-world class imbalance, where high-risk but clinically critical diseases occur infrequently compared to benign cases (Fig.~\ref{fig:isic}(a)). This imbalance reflects practical screening scenarios and motivates uncertainty-aware inference for safety-critical decisions.

For a fair comparison, the same network backbone is used across all experiments. A pretrained ResNet-50 serves as a fixed feature extractor, followed by three FC layers implemented in hardware. Three weight representations are evaluated for the FC layers: fixed-point deterministic weights (NN), uni-modal Gaussian weights (BNN), and MoG weights (MoG BNN). The MoG BNN is evaluated at $K=3$, balancing distributional expressiveness with area-normalized throughput. Probabilistic models are evaluated using 20 samples per input to estimate predictive uncertainty. The selection of these two operating points are further discussed below.

\subsection{Rare-Disease Screening}

\begin{figure}[t!]
    \centering
    \vspace{-0.6cm}
    \includegraphics[width=\columnwidth]{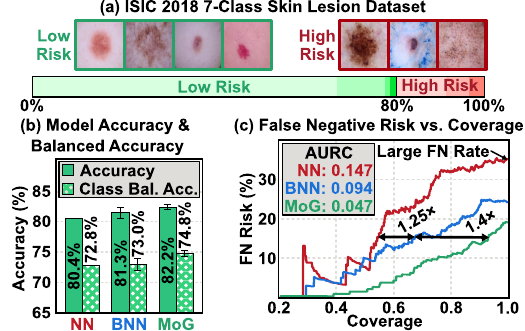}
    \caption{(a) The ISIC 2018 Task 3 dataset~\cite{ISIC1,ISIC2}, suffers from extreme real-world imbalance between high and low risk classes. MoG BNN expressiveness offers improved (b) class balanced accuracy and (c) false negative classification risk.}
    \label{fig:isic}
    \vspace{-0.3cm}
\end{figure}

Rare-disease screening performance is evaluated using class-balanced accuracy (the primary metric for imbalanced datasets) and risk–coverage analysis, emphasizing correct detection of minority, high-risk classes. Compared to deterministic and uni-modal Bayesian baselines, MoG BNNs improve balanced class accuracy by $\geq$1.8\%, achieving 82.2\% overall accuracy and 74.8\% balanced accuracy (Fig.~\ref{fig:isic}(b)). This improvement demonstrates that multi-modal weights better capture heterogeneous decision boundaries induced by imbalanced medical datasets, without degrading majority-class performance.

To assess classification safety, uncertainty-based selective prediction is employed, where high-entropy predictions are deferred for expert review or rescreening. At fixed false-negative risk (missed cancerous classification), MoG BNNs increase model coverage by 1.4$\times$ relative to uni-modal BNNs and reduce the area under risk-coverage curve AURC by 2$\times$ (Fig.~\ref{fig:isic}(c)). These results confirm that increased expressiveness directly enhances the model’s ability to trade coverage for risk, an essential property for medical screening.

Table~\ref{tab:comparison} also reports the accuracy and AURC improvements with CIFAR-10 to further validate consistent MoG BNN gains across datasets. 

\begin{figure}[b!]
    \centering
    \vspace{-0.3cm}
    \includegraphics[width=\columnwidth]{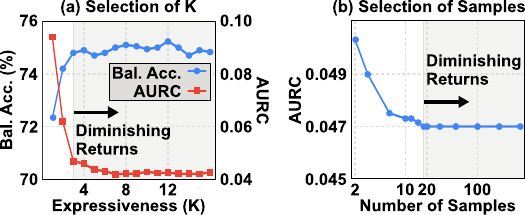}
    \caption{(a) Balanced accuracy and uncertainty awareness (area under risk-coverage curve (AURC)) across different expressiveness levels with diminishing returns beyond $K = 3$. (b) AURC across different number of samples per input for uncertainty quantification (evaluated at $K = 3$).}
    \label{fig:Kchoice}
\end{figure}

As shown in Figure~\ref{fig:Kchoice}(a), $K = 3$ was selected for analysis of the MoG BNN as it represents the inflection point of diminishing returns for both balanced accuracy and uncertainty quantification (AURC). While the $K = 1$--$16$ programmability maintains deployment flexibility for more complex applications, $K = 3$ provides the optimal balance between predictive performance and parameter storage overhead for the targeted task. For uncertainty estimation, 20 inference samples per input was identified as the threshold beyond which the uncertainty estimates achieved stability when $K = 3$ (Fig.~\ref{fig:Kchoice}(b)).

\subsection{At-Home Screening Robustness}

\begin{figure}[t!]
    \centering
    \includegraphics[width=\columnwidth]{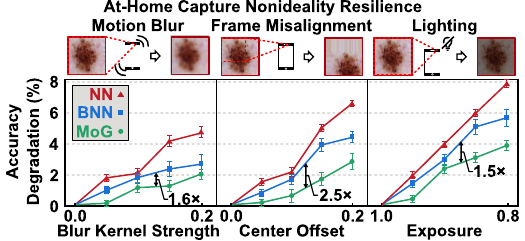}
    \caption{At-home screening offers rapid results and enhances user privacy, but suffers from user error. MoG BNNs further improve the resilience to these non-idealities.}
    \label{fig:home}
    \vspace{-0.3cm}
\end{figure}

At-home screening introduces uncontrolled acquisition conditions, including motion blur, focus misalignment, and variation in exposure. To evaluate robustness, these nonidealities are synthetically injected into the ISIC test set using image perturbations (Fig.~\ref{fig:home}). Motion blur is modeled by convolving each image with a spatial kernel that blends neighboring pixels, where the kernel strength controls the contribution intensity of surrounding pixels and thus the blur severity. Under this distortion, the MoG BNN demonstrates an average $1.6\times$ improvement in accuracy retention compared to a uni-modal BNN. 

Focus misalignment is emulated by translating the image frame such that the lesion is intentionally shifted away from the center of view; the perturbation strength is defined by the pixel offset magnitude from the nominal centered position. This condition reflects user-induced framing errors during handheld capture and yields an average $2.5\times$ robustness improvement for the MoG BNN. 

Lighting variation is introduced by uniformly reducing image exposure to simulate dim acquisition environments, producing lower-contrast photographs typical of non-clinical settings. In this scenario, the MoG BNN provides an average $1.5\times$ improvement in resilience. 

Across all perturbations, MoG BNNs exhibit the smallest degradation in classification accuracy as distortion severity increases. This resilience arises from the inherent BNN robustness to noisy inputs and the ability of multi-modal weight distributions to capture distinct acquisition regimes, preventing overconfident predictions under degraded inputs. The results highlight the importance of expressive uncertainty modeling for reliable inference under user-induced variability typical of at-home medical screening.

%% file: 03_Hardware.tex
\begin{figure}[b!]
    \centering
    \vspace{-0.4cm}
    \includegraphics[width=\columnwidth]{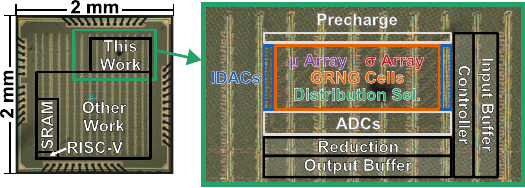}
    \caption{Annotated die photo of this work's MoG BNN hardware.}
    \label{fig:chip}
\end{figure}

\section{Chip Validation}
\label{sec:hardware}

A prototype chip fabricated on a commercial 65~\qty{}{\nm} PDK (Fig.~\ref{fig:chip}) provides validation measurements for this system's MoG Bayesian FC layers.

\subsection{Area and Energy Breakdown}

\begin{figure}[t!]
    \centering
    \includegraphics[width=\columnwidth]{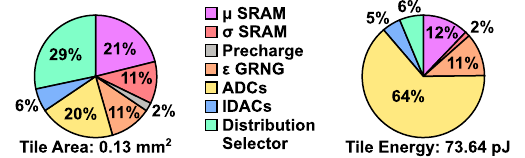}
    \caption{Area and energy breakdown of the $64\times8$ prototype MoG BNN tile.}
    \label{fig:area}
    \vspace{-0.3cm}
\end{figure}

Figure~\ref{fig:area} breaks down the area and energy of this work. The compute tile occupies 0.13~\qty{}{\mm\squared} and with the associated controller consumes 0.225~\qty{}{\mm\squared}. The tile requires 73.64~\qty{}{\pico\joule} per MVM operation when operating at $V_{DD} =$ 1~\qty{}{\volt}. ADCs account for 64\% of the total MVM energy, underscoring that the integrated in-word GRNG and distribution selectors incur relatively low overhead. Furthermore, the RISC-V core and associated SRAM requires 3 clock cycles and 59.3~\qty{}{\pico\joule} for both read and write logic.

This prototype design showcases a $64\times8$ word tile, where more chip area or tile reuse is required for larger networks. The 3-layer FC network used for the skin lesion evaluation in Section~\ref{sec:model} validates this scaling. Following feature extraction, the MoG BNN FC layer execution—considering tile reuse, $K = 3$, and 20 samples per input—for the skin lesion screening application consumes 9.484~\qty{}{\micro\joule} per inference. When accounting for the system-level overhead of the RISC-V core for instruction/data management, the total post-feature-extraction energy is 17.502~\qty{}{\micro\joule} with a latency of 72.15~\qty{}{\ms} per 20-sample inference, confirming the practicality for real-time at-home screening applications. This latency also decreases as the hardware tile size increases proportionally.

Feature extraction is performed prior to this work's contribution using a ResNet-50 backbone deployed on a local neural processing unit (NPU). On commercially available edge platforms such as the Snapdragon 8 Gen 3, NPU-accelerated ResNet-50 inference requires $\sim$1.5~\qty{}{\ms} latency~\cite{qualcom}. Using a conservative $3$~\qty{}{\watt} power draw, this corresponds to $\sim$4.5~\qty{}{\milli\joule} per extracted feature image, highlighting the low latency pipeline of feature extraction and the energy efficiency of the proposed MoG BNN FC layer hardware for risk-aware screening.

\subsection{GRNG Bias Tuning and System-Level Tradeoffs}

\begin{figure}[t!]
    \centering
    \includegraphics[width=\columnwidth]{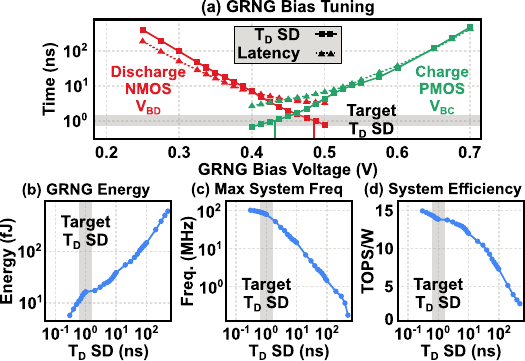}
    \caption{(a) Tuning the GRNG bias voltages ($V_{BC}$ and $V_{BD}$) impacts the sample's latency and time-pulse ($T_D$) standard deviation (SD). This impacts the (b) GRNG sample energy, (c) maximum system operating frequency, and (d) system's network efficiency (TOPS/\qty{}{\watt}).}
    \label{fig:grng_tuning}
    \vspace{-0.2cm}
\end{figure}

System operating frequency is bounded by the latency and pulse width ($T_D$) of the GRNG cells, as each inference cycle must allow sufficient time to fully charge and discharge the sampling capacitors to produce statistically valid Gaussian samples. Consequently, the GRNG defines the critical path of the chip. Its behavior is governed by the bias voltages $V_{BC}$ and $V_{BD}$, which control the PMOS charging and NMOS discharging phases, respectively, and thereby determine both the overall latency and the $T_D$ standard deviation (SD) (Fig.~\ref{fig:grng_tuning}(a)).

The chosen $T_D$ SD establishes the operating point of the GRNG and directly impacts the energy per sample (Fig.~\ref{fig:grng_tuning}(b)), achievable clock frequency (Fig.~\ref{fig:grng_tuning}(c)), and throughput (Fig.~\ref{fig:grng_tuning}(d)). Targeting a $T_D$ SD of 1~\qty{}{\ns} provides sufficient $\sigma\epsilon$ dynamic range on the bitlines while maximizing energy efficiency, such that when multiplied by the minimum non-zero input ($x=1$), the bitline’s discharge between $0$ and $1$ SD can be read with at least 1 least significant bit of the ADCs precision. At this operating point, the chip achieves 16.3~\qty{}{\femto\joule} per GRNG sample, a 74.1~\qty{}{\mega\hertz} system clock, and 13.9 TOPS/\qty{}{\watt} efficiency. Note that the max frequency considers the full charge and discharge time of the GRNG's capacitors, such that $> 99.7\%$ of samples complete full integration.

\begin{figure}[b!]
    \centering
    \vspace{-0.2cm}
    \includegraphics[width=\columnwidth]{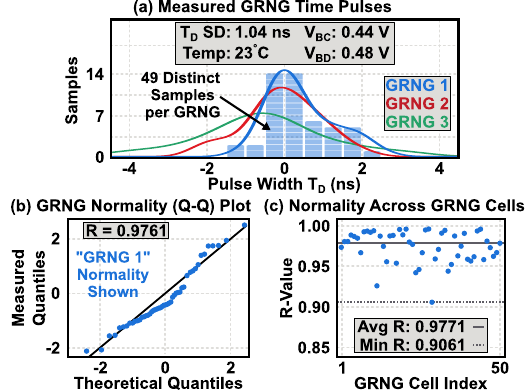}
    \caption{(a) GRNG time pulse samples (49 distinct points per GRNG per clock edge). (b) GRNG normality (Q-Q) plot for a single GRNG cell. (c) Normality correlation coefficient across all measured dies.}
    \label{fig:grng_results}
\end{figure}

\subsection{GRNG Sample Quality and Gaussian Validation}

The GRNG cells generate 49 discrete charge/discharge samples per inference cycle. Measured time-domain pulses and the associated distributions of multiple GRNG cells are shown in Figure~\ref{fig:grng_results}(a). Their statistical quality is also validated both algorithmically (Section~\ref{sec:model}) and empirically via normality probability (Q--Q) plots across 50 dies. The samples exhibit strong Gaussian behavior, with a mean correlation coefficient $R = 0.9771$. Across the measured population, 90\% of cells achieve $R > 0.95$, with a minimum observed value of 0.9061 (Fig.~\ref{fig:grng_results}(b,c)), confirming suitability for quantized inference.

\subsection{Throughput and Scalable-$K$ Tradeoffs}

Supporting programmable $K$ introduces a modest area overhead relative to fixed-$K$ designs, which require only $K-1$ distribution selectors per word group. In this chip's implementation, the overhead is determined by the tile organization, where each column can process $\lfloor 64/K \rfloor$ weights per cycle from a 64-row array. Consequently, the fraction of the reported distribution selector circuit overhead that would otherwise be unnecessary in a fixed-$K$ design scales according to $\frac{64-\lfloor 64/K \rfloor (K-1)}{64}$. This additional circuitry enables post-fabrication tuning of model expressiveness versus area-normalized throughput, as well as the ability to introduce new mixture modes during deployment—capabilities not possible in a fixed-$K$ architecture. Rather than locking the chip to a single operating point, the same silicon can be repurposed across applications with different uncertainty and performance requirements.

Area-normalized throughput also correspondingly scales with $K$, reflecting the reduced number of weights processed per cycle on a fixed-size tile as the mixture order increases. While the memory footprint grows linearly with $K$ because each component must be stored, the energy remains nearly constant since only one distribution is sampled in a given inference cycle and the remaining components remain inactive.

\subsection{PVT Robustness and Environmental Sensitivity}

\begin{figure}[b!]
    \centering
    \vspace{-0.2cm}
    \includegraphics[width=\columnwidth]{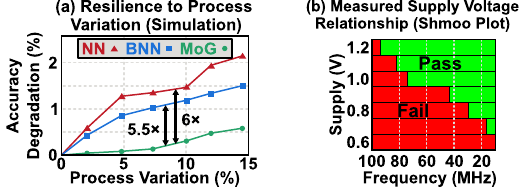}
    \caption{(a) Simulated model resilience to increasing process variation within the CIM tile MAC operations. (b) Shmoo plot relationship between supply voltage and the system's maximum operating frequency.}
    \label{fig:pvt}
\end{figure}

Process variation among the CIM tile was simulated via injecting noise in the weights. BNNs are inherently tolerant to such noise sources~\cite{pei_ICCAD24}, and the MoG BNN expressiveness exhibited further resilience with $5.5\times$ lower skin lesion classification degradation (Fig.~\ref{fig:pvt}(a)).

The Shmoo plot in Fig.~\ref{fig:pvt}(b) illustrates the voltage--frequency tradeoff needed to maintain the target 1~ns (or minimum possible) $T_{D}$ SD, while Table~\ref{tab:grng_supply} confirms stable statistical quality ($R \ge 0.974$) across the supply range. Notably, achieving 1~ns $T_{D}$ SD requires $V_{DD} \ge 1.0$~V; below this threshold, reduced current drive limits charge/discharge speeds, causing the minimum achievable SD to exceed 1~ns.

\begin{table}[t!]
\centering
\caption{Measured GRNG Supply Voltage Stability}
\label{tab:grng_supply}
\renewcommand{\arraystretch}{1.15}
\begin{tabular*}{\linewidth}{@{\extracolsep{\fill}}ccccc}
\hline
\makecell{\textbf{$V_{\text{DD}}$}\\(V)} &
\makecell{\textbf{Avg. Q--Q}\\\textbf{R-value $\uparrow$}} &
\makecell{\textbf{$T_\text{D}$ SD}\\(ns)} &
\makecell{\textbf{Avg. Latency}\\(ns)} &
\makecell{\textbf{Max Freq.}\\\textbf{(MHz)}} \\
\hline
1.2 & 0.9740 & 1.050  & 2.37  & 93.22 \\
1.1 & 0.9747 & 1.037  & 2.612 & 81.96 \\
1.0 & 0.9771 & 1.040  & 3.430 & 74.10 \\
0.9 & 0.9759 & 1.712  & 5.250 & 43.66 \\
0.8 & 0.9757 & 5.249  & 13.790 & 30.36 \\
0.7 & 0.9756 & 17.170 & 40.750 & 16.72 \\
\hline
\end{tabular*}
\end{table}

Similarly, temperature sweeps from 23~\qty{}{\celsius} to 60~\qty{}{\celsius} reduce $T_{D}$ standard deviation and GRNG latency by approximately 1.5$\times$ and 1.1$\times$, respectively, while maintaining high statistical fidelity (Table~\ref{tab:grng_temp}). Since the zero-mean property is intrinsic to the differential structure, it persists across temperature and voltage variations without requiring static D2D offset recalibration. Although environmental fluctuations do influence the global timing spread, these shifts are effectively mitigated by tuning the global bias voltages ($V_{BC}$ and $V_{BD}$). Crucially, this system-level tuning preserves the target operating point of a 1~\qty{}{\ns} $T_{D}$ SD without compromising the inherently calibration-free, zero-mean quality of the generated samples.
\begin{table}[t!]
\centering
\caption{Measured GRNG Temperature Stability}
\label{tab:grng_temp}
\renewcommand{\arraystretch}{1.15}
\begin{tabular*}{\linewidth}{@{\extracolsep{\fill}}cccc}
\hline
\makecell{\textbf{Temperature}\\($^\circ$C)} &
\makecell{\textbf{Avg. Q--Q}\\\textbf{R-value $\uparrow$}} &
\makecell{\textbf{$T_\text{D}$ SD}\\(ns)} &
\makecell{\textbf{Avg. Latency}\\(ns)} \\
\hline
23 & 0.9771 & 1.04 & 3.43 \\
30 & 0.9873 & 1.02 & 3.25 \\
40 & 0.9794 & 0.98 & 3.19 \\
50 & 0.9803 & 0.80 & 3.17 \\
60 & 0.9798 & 0.72 & 3.09 \\
\hline
\end{tabular*}
\vspace{-0.3cm}
\end{table}

\subsection{Impact of Non-Idealities on Entropy Stability} 
To evaluate the proposed GRNG's robustness against dynamic and aging-induced variations, a comprehensive sensitivity analysis was performed. 
\begin{table}[b!]
\centering
\vspace{-0.3cm}
\caption{Average simulated GRNG $R$-value across non-idealities}
\label{tab:nonideal}
\renewcommand{\arraystretch}{1.15}
\begin{tabular*}{\linewidth}{@{\extracolsep{\fill}}ccc}
\hline
\multirow{2}{*}{\makecell{\textbf{Variation}\\\textbf{Strength (\%)}}} & \multicolumn{2}{c}{\textbf{Non-Idealities}} \\ 
\cline{2-3}
& \makecell{\textbf{Thermal Noise}} & \makecell{\textbf{Aging (BTI)}} \\
\hline
0 (Measured) & 0.9771 & \textbf{0.9771} \\
2.5 & 0.9774 & 0.9764 \\
5.0 & 0.9779 & 0.9758 \\
7.5 & 0.9784 & 0.9753 \\
10.0 & \textbf{0.9789} & 0.9749 \\
\hline
\end{tabular*}
\end{table}

\subsubsection{Thermal Noise}
While the GRNG leverages static process variation for entropy, dynamic thermal noise introduces marginal timing jitter. Since static variation inherently dominates thermal noise in near-threshold regimes~\cite{thermal}, additive thermal jitter (simulated up to 10\% of the process variation standard deviation, superimposed on measured data) does not degrade the Gaussian $R$-value. Instead, this uncorrelated zero-mean noise strictly perturbates static samples, slightly improving $R$ from $0.9771$ to $0.9789$ (Table~\ref{tab:nonideal}).

\subsubsection{Leakage}
Sub-threshold leakage is exponentially dependent on localized $V_{th}$ variations~\cite{leakage1}. Consequently, baseline leakage disparities are natively absorbed into the measured process variation profile. Furthermore, dynamic leakage surges—such as those induced by elevated temperatures~\cite{leakage2}—are intrinsically captured within the aforementioned temperature variation measurements.

\subsubsection{Power Cycles}
Power cycling induces transient global supply rail settlements and voltage droops~\cite{power1, power3}. The differential topology explicitly rejects these global shifts as common-mode, preserving the original distribution. Residual localized effects, such as random telegraph noise or transient charge trapping~\cite{power2}, manifest as independent device-level fluctuations that can thus be modeled similar to thermal noise.

\subsubsection{Aging}
Device aging induces systematic threshold voltage drift; however, low-voltage operation substantially mitigates bias temperature instability (BTI), hot carrier injection, and time-dependent dielectric breakdown~\cite{aging1,aging2}. BTI remains the dominant degradation mechanism~\cite{aging2,aging3}, increasing $|V_{th}|$ via a sublinear power-law ($\propto \text{time}^n, n<1$) proportional to initial $V_{th}$~\cite{aging4}. Consequently, BTI causes a global positive mean $V_{th}$ shift and modest variance expansion, non-linearly expanding the discharge times of slower devices more aggressively. To simulate this aging trend, the variation percentages in Table~\ref{tab:nonideal} represent the maximum applied shift to the underlying threshold voltage (up to 10\%). Across these shifts, the resulting tail expansion minimally degrades Gaussian quality, dropping $R$ from $0.9771$ to $0.9749$. Crucially, GRNG transistors remain unbiased for $\sim$93\% of the operating cycle, enabling substantial BTI recovery and decelerating degradation.

Similar to temperature or voltage fluctuations, systematic variance shifts induced by aging or other non-idealities can be compensated through periodic global bias adjustments to maintain a constant delay standard deviation (e.g., 1~ns $T_D$ SD). Regardless, such shifts act as common-mode noise between the symmetric entropy banks, inherently preserving the calibration-free zero-mean characteristics.

%% file: 05_Conclusions.tex
\begin{table*}[h!]
\centering
\caption{Comparison to Other Work}
\label{tab:comparison}
\renewcommand{\arraystretch}{1.2}
\begin{tabular}{lccccc}
\hline
\rowcolor{gray!25}
\multicolumn{6}{l}{\textbf{Hardware Evaluation}} \\
\hline
 & \textbf{\textcolor{green!60!black}{This Work}} 
 & \textbf{ISSCC~25~\cite{liu_ISSCC25}} 
 & \textbf{ISSCC~25~\cite{STT_ISSCC25}} 
 & \textbf{JSSC~23~\cite{intel_JSSC23}} 
 & \textbf{ICCAD~24~\cite{pei_ICCAD24}} \\
\hline

\textbf{Implementation} & ASIC & ASIC & ASIC & ASIC & Simulation \\
\arrayrulecolor{gray!30}\hline\arrayrulecolor{black}

\textbf{Technology Node [\qty{}{\mathbf{\nm}}]} & 65 & 65 & 22 & 22 & 65 \\
\arrayrulecolor{gray!30}\hline\arrayrulecolor{black}

\textbf{RNG Source} & Process Variation & Thermal & STT-MRAM & TI-Hadamard & V$_{\text{DD}}$ Variation \\
\arrayrulecolor{gray!30}\hline\arrayrulecolor{black}

\textbf{Precision} & INT 8/4 & INT 8/4 & INT 8 & INT/BF/FP 8/16/32 & INT 8/4 \\
\arrayrulecolor{gray!30}\hline\arrayrulecolor{black}

\textbf{Area [\qty{}{\mathbf{\mm\squared}}]} & 0.225 & 0.816 & 3.49 & 3.88 & -- \\
\arrayrulecolor{gray!30}\hline\arrayrulecolor{black}

\textbf{In-Word GRNG} 
& \textcolor{green!60!black}{\checkmark} 
& \textcolor{green!60!black}{\checkmark} 
& \textcolor{red}{\texttimes} 
& \textcolor{red}{\texttimes} 
& \textcolor{red}{\texttimes} \\
\arrayrulecolor{gray!30}\hline\arrayrulecolor{black}

\textbf{Calibration-Free GRNG} 
& \textcolor{green!60!black}{\checkmark} 
& \textcolor{red}{\texttimes} 
& \textcolor{red}{\texttimes} 
& \textcolor{green!60!black}{\checkmark} 
& \textcolor{red}{\texttimes} \\
\arrayrulecolor{gray!30}\hline\arrayrulecolor{black}

\textbf{Uncertainty-Awareness} 
& \textcolor{green!60!black}{\checkmark} 
& \textcolor{green!60!black}{\checkmark} 
& \textcolor{red}{\texttimes} 
& \textcolor{green!60!black}{\checkmark} 
& \textcolor{green!60!black}{\checkmark} \\
\arrayrulecolor{gray!30}\hline\arrayrulecolor{black}

\textbf{RNG Throughput [GSa/s]} & \textbf{37.9} & 5.12 & -- & 4.65--7.31 & -- \\
\arrayrulecolor{gray!30}\hline\arrayrulecolor{black}

\begin{tabular}{@{}l@{}}\textbf{Normalized RNG Throughput} \\ \textbf{[GSa/s/\qty{}{\mathbf{\mm\squared}}]}\end{tabular} 
& \textbf{168.6 (923)$^\ast$ }
& 11.4 (62.3)$^\ast$ 
& -- 
& 1.2--1.88 
& -- \\
\arrayrulecolor{gray!30}\hline\arrayrulecolor{black}

\textbf{RNG Efficiency [fJ/Sa]} & \textbf{16.3} & 360 & -- & 1080--1690 & 102 \\
\arrayrulecolor{gray!30}\hline\arrayrulecolor{black}

\begin{tabular}{@{}l@{}}\textbf{Normalized FC-Layer NN} \\ \textbf{Throughput [GOp/s/\qty{}{\mathbf{\mm\squared}}]}\end{tabular} 
& \textbf{337.3 (1847)$^\ast$ }
& -- 
& 36.7 
& 309--515 
& -- \\

\hline
\rowcolor{gray!25}
\multicolumn{6}{l}{\textbf{Algorithmic Evaluation}} \\
\hline

\textbf{Expressiveness} 
& \textbf{MoG} ($K=3$ Shown)
& \multicolumn{4}{c}{Uni-modal Gaussian} \\
\arrayrulecolor{gray!30}\hline\arrayrulecolor{black}

\textbf{Class-Balanced Accuracy} 
& \textbf{74.8\%$^\dagger$ \quad 88.1\%$^\ddagger$ }
& \multicolumn{4}{c}{73.0\%$^\dagger$ \quad 85.6\%$^\ddagger$} \\
\arrayrulecolor{gray!30}\hline\arrayrulecolor{black}

\begin{tabular}{@{}l@{}}\textbf{At-Home Screening} \\ \textbf{Accuracy Degradation $\downarrow$}\end{tabular}
& \textbf{2~--~4\%$^\dagger$ }
& \multicolumn{4}{c}{2.5~--~6\%$^\dagger$} \\
\arrayrulecolor{gray!30}\hline\arrayrulecolor{black}

\textbf{Area-Under-Risk-Coverage-Curve $\downarrow$} 
& \textbf{0.047$^\dagger$ \quad 0.028$^\ddagger$ }
& \multicolumn{4}{c}{0.094$^\dagger$ \quad 0.036$^\ddagger$} \\

\hline
\end{tabular}

\vspace{2pt}
\footnotesize
\begin{minipage}{\linewidth}
\raggedright
$~~~~~^\ast$Scaled to 22\,nm based on predictive scaling abstraction from~\cite{scaling}\\
$~~~~~^\dagger$ISIC 2018 (ResNet-50 Feature Extractor) \\
$~~~~~^\ddagger$CIFAR-10 (MobileNet Feature Extractor)
\end{minipage}
\end{table*}

\section{Conclusions}

This work presents the first CIM hardware platform enabling MoG BNN inference, providing expressive and uncertainty-aware computation at energy efficiencies suitable for edge and at-home medical screening. Fabricated in 65~\qty{}{\nm} CMOS, the chip operates at 1~\qty{}{\volt} and 74~\qty{}{\mega\hertz} with a calibrated $T_D$ standard deviation of 1~\qty{}{\ns}. Table~\ref{tab:comparison} provides a summary of key performance metrics compared to SOTA BNN accelerators and their associated uni-modal BNN limitations. As the first MoG BNN hardware targeting medical imaging dataset, we isolate our algorithmic advantages by assuming all prior SOTA architectures can perfectly execute ideal, quantized uni-modal BNNs without device variation. This frames the algorithmic comparison purely around the performance gap dictated by distribution expressiveness (MoG vs. uni-modal). To ensure broad comparability alongside the ISIC 2018 dataset, we also evaluate on the standard CIFAR-10 benchmark. Compared to SOTA in-word GRNG~\cite{liu_ISSCC25}, this design’s process variation not only exploits calibration-free deployment, but also outweighs thermal noise, enabling faster clocks and reduced inverter short-circuit current. This lowers energy per sample by $22\times$ (16.3~\qty{}{\femto\joule}/sample) and increases parallel GRNG throughput by $7.4\times$ (37.9\,GSa/s) ($14.8\times$ area-normalized: 168.6\,GSa/s/\qty{}{\mm\squared}). The lower GRNG latency allows this design to also merge $\mu$ and $\sigma$ into a single word with distribution selectors for MoG sampling, offering two key advantages: (1) merged $\mu$ and $\sigma$ tiles allow one distribution selector per word, reducing selector overhead by $2\times$, and (2) faster $\sigma\epsilon$ computation improves uncertainty quantification across repeated sampling, reducing inference latency proportional to the $7.4\times$ throughput gain.

Compared to uni-modal BNN designs, the proposed MoG BNN provides enhanced uncertainty awareness by explicitly preserving multiple parameter hypotheses during inference. In contrast, some SOTA BNN accelerators prioritize noise resilience~\cite{STT_ISSCC25} and reuse GRNG samples after a single generation cycle, trading uncertainty awareness for reduced sampling cost. Algorithmic evaluation on the ISIC 2018 skin lesion dataset shows that hardware support for MoG Bayesian inference directly improves the safety of at-home screening by better representing dataset imbalance and enabling more reliable risk-aware predictions. Relative to SOTA uni-modal BNN models, the MoG BNN achieves up to 1.8\% higher class-balanced accuracy, 1.4$\times$ increased model coverage at fixed false-negative risk, $>$1.5$\times$ improved resilience under at-home degradation, and 5.5$\times$ greater robustness to CIM noise and process variation. These gains are achieved without sacrificing energy efficiency, validating the practicality of expressive Bayesian inference in memory. Overall, this work establishes a scalable and energy-efficient foundation for MoG Bayesian inference, enabling safer, more reliable, and privacy-preserving edge AI with flexibility for at-home screening tasks.